\documentclass[prl,twocolumn,showpacs,superscriptaddress,floatfix]{revtex4}
 
\usepackage{graphics}  
\usepackage{makeidx}
\usepackage{colordvi}
\RequirePackage{xspace}

\usepackage{relsize}
\def\babar{\mbox{\slshape B\kern-0.1em{\smaller A}\kern-0.1em
    B\kern-0.1em{\smaller A\kern-0.2em R}}}
\def\epem       {\ensuremath{e^+e^-}\xspace}
\def\qqbar {\ensuremath{q\overline q}\xspace}
\def\ccbar {\ensuremath{c\overline c}\xspace}
\def\piz   {\ensuremath{\pi^0}\xspace}
\def\pip   {\ensuremath{\pi^+}\xspace}
\def\pim   {\ensuremath{\pi^-}\xspace}

\def\Kbar  {\kern 0.2em\overline{\kern -0.2em K}{}\xspace}

\def\Km    {\ensuremath{K^-}\xspace}
\def\KS    {\ensuremath{K^0_{\scriptscriptstyle S}}\xspace}
\def\Dbar    {\kern 0.2em\overline{\kern -0.2em D}{}\xspace}

\def\Dz      {\ensuremath{D^0}\xspace}
\def\Dzb     {\ensuremath{\Dbar^0}\xspace}
\def\Dp      {\ensuremath{D^+}\xspace}
\def\Dm      {\ensuremath{D^-}\xspace}

\def\Dstarzb {\ensuremath{\Dbar^{*0}}\xspace}
\def\Dstarp  {\ensuremath{D^{*+}}\xspace}
\def\Dstarm  {\ensuremath{D^{*-}}\xspace}
\def\Dstarpm {\ensuremath{D^{*\pm}}\xspace}

\def\Bbar    {\kern 0.18em\overline{\kern -0.18em B}{}\xspace}

\def\BB      {\ensuremath{B\Bbar}\xspace}
\def\Bz      {\ensuremath{B^0}\xspace}
\def\Bzb     {\ensuremath{\Bbar^0}\xspace}
\def\BzBzb   {\ensuremath{\Bz {\kern -0.16em \Bzb}}\xspace}
\def\Bu      {\ensuremath{B^+}\xspace}

\def\Bp      {\ensuremath{\Bu}\xspace}

\def\jpsi     {\ensuremath{{J\mskip -3mu/\mskip -2mu\psi\mskip 2mu}}\xspace}
\def\Y#1S{\ensuremath{\Upsilon{(#1S)}}\xspace}% no space before {...}!
\def\FourS {\Y4S}
\def\bpsiks     {\ensuremath{\Bz \to \jpsi \KS}\xspace}

\def\Bztodstdstks {\ensuremath{\Bz \to \Dstarp \Dstarm \KS}\xspace}

\def\upsbb   {\ensuremath{\FourS \to \BB}\xspace}
\def\mes        {\mbox{$m_{\rm ES}$}\xspace}
\newcommand{\mev}{\ensuremath{\mathrm{\,Me\kern -0.1em V}}\xspace}
\newcommand{\gevc}{\ensuremath{{\mathrm{\,Ge\kern -0.1em V\!/}c}}\xspace}
\newcommand{\mevc}{\ensuremath{{\mathrm{\,Me\kern -0.1em V\!/}c}}\xspace}
\newcommand{\gevcc}{\ensuremath{{\mathrm{\,Ge\kern -0.1em V\!/}c^2}}\xspace}
\newcommand{\mevcc}{\ensuremath{{\mathrm{\,Me\kern -0.1em V\!/}c^2}}\xspace}
\def\to                 {\ensuremath{\rightarrow}\xspace}
\newcommand{\stat}{\ensuremath{\mathrm{(stat)}}\xspace}
\newcommand{\syst}{\ensuremath{\mathrm{(syst)}}\xspace}
\def\pep2{PEP-II}

\def\CP                {\ensuremath{C\!P}\xspace}
\def\stwob{\ensuremath{\sin\! 2 \beta   }\xspace}
\def\ctwob{\ensuremath{\cos\! 2 \beta   }\xspace}
\def\deltat{\ensuremath{{\rm \Delta}t}\xspace}
\def\deltamd{\ensuremath{{\rm \Delta}m_d}\xspace}
\newcommand{\progtp}    [1]  {{Prog.\ Theor.\ Phys.\ {\bf #1}}}

\begin{document}  

\begin{flushleft}
\babar-PUB-06/023\\
SLAC-PUB-12045\\
%hep-ex/0608016
\end{flushleft}

\title{
{\large  \boldmath
Measurement of the Branching Fraction and Time-Dependent  \\
\CP\ Asymmetry in the Decay $\Bztodstdstks$}
}
 
%% author list as of 01-Jul-2006 (596 authors)
%
\author{B.~Aubert}
\author{M.~Bona}
\author{D.~Boutigny}
\author{F.~Couderc}
\author{Y.~Karyotakis}
\author{J.~P.~Lees}
\author{V.~Poireau}
\author{V.~Tisserand}
\author{A.~Zghiche}
\affiliation{Laboratoire de Physique des Particules, IN2P3/CNRS et Universit\'e de Savoie,
 F-74941 Annecy-Le-Vieux, France }
\author{E.~Grauges}
\affiliation{Universitat de Barcelona, Facultat de Fisica, Departament ECM, E-08028 Barcelona, Spain }
% ...
\collaboration{The \babar\ Collaboration}
\noaffiliation

\begin{abstract}
We study the decay $\Bztodstdstks$ using
$(230\pm 2) \times 10^{6} \BB$ pairs collected
by the \babar\  detector at the PEP-II $B$ factory.
We measure a branching fraction 
$\mathcal{B}(\Bztodstdstks)=(4.4\pm0.4\pm0.7)\times 10^{-3}$
and find  evidence for the decay 
$\Bz\to\Dstarm D^+_{s1}(2536)$ with a significance 
of $4.6\,\sigma$. A time-dependent \CP asymmetry analysis is
also performed to study the possible resonant contributions
to $\Bztodstdstks$
and the sign of \ctwob . Our measurement indicates that 
there is a sizable resonant contribution to the decay 
$\Bz\to\Dstarp\Dstarm\KS$ from a unknown $D^+_{s1}$ state with large
width, and that \ctwob is positive at the 94\,\% confidence level
under certain theoretical assumptions.

\end{abstract}
 
% PACS, the Physics and Astronomy Classification Scheme.
  
\maketitle    

In the standard model framework, 
\CP violation arises from
a complex phase in the Cabibbo-Kobayashi-Maskawa (CKM) quark-mixing
matrix~\cite{CKM}. Measurements of \CP asymmetries by the
\babar~\cite{Aubert:2002ic} and Belle~\cite{Abe:2002px}
collaborations have firmly established this effect in the decay
\bpsiks~\cite{conjugate} and related modes that are governed by
the $b\to\ccbar s$ transition.
Since both $\Bz$ and $\Bzb$ mesons
can decay to the final state $\Dstarp\Dstarm\KS$ and this process 
is dominated by a single weak phase, $W$-emission $b\to\ccbar s$ transition,
a time-dependent \CP violating asymmetry is expected.

In the approximation of neglecting penguin contributions for the
decay $\Bz\to\Dstarp\Dstarm\KS$, there is no direct \CP violation.
The time-dependent decay rate asymmetry of $\Bz\to\Dstarp\Dstarm\KS$ in 
the half Dalitz space $s^+\le s^-$ 
or $s^+\ge s^-$ can be written as~\cite{Browder:1999ng}
\begin{eqnarray}
A(t)&\equiv&\frac{\Gamma_{\Bzb}-\Gamma_{\Bz}}{\Gamma_{\Bzb}+\Gamma_{\Bz}}
=\eta_y\frac{J_c}{J_0}\cos(\Delta m_dt)- \nonumber\\
&&\left(\frac{2J_{s1}}{J_0}\stwob +\eta_y \frac{2J_{s2}}{J_0}\ctwob\right)
\sin(\Delta m_dt),
\end{eqnarray}
where $s^+\equiv m^2(\Dstarp\KS)$ and $s^-\equiv m^2(\Dstarm\KS)$, 
$\Gamma_{\Bz}$ ($\Gamma_{\Bzb}$) is the decay rate for \Bz (\Bzb) 
to $\Dstarp\Dstarm\KS$ at a proper time $t$ after production, $\Delta m_d$ is the
mass difference between the two $\Bz$ mass eigenstates, and 
$\eta_y=-1(+1)$ for $s^+\le s^- (s^+\ge s^-)$. 
The parameters $J_0$,$J_c$,$J_{s1}$ and $J_{s2}$ are the
integrals over the half Dalitz phase space with $s^+<s^-$ of the functions
$|a|^2+|\bar{a}|^2$, $|a|^2-|\bar{a}|^2$, 
$\mathop{\cal R\mkern -2.0mu\mit e}(\bar{a}a^*)$ and
$\mathop{\cal I\mkern -2.0mu\mit m}(\bar{a}a^*)$, 
where $a$ and $\bar{a}$ are the decay amplitudes of
$\Bz\to\Dstarp\Dstarm\KS$ and $\Bzb\to\Dstarp\Dstarm\KS$,
respectively.

If the decay $\Bz\to\Dstarp\Dstarm\KS$ has only a non-resonant component,
the parameter $J_{s2}=0$ and $J_c$ is at the few percent level~\cite{Browder:1999ng}. 
The \CP\ asymmetry can be extracted by fitting the
\Bz time-dependent decay distribution. The measured \CP\ asymmetry is 
\stwob multiplied by a factor of $2J_{s1}/J_0$ because the final state is an 
admixture of \CP\ eigenstates with different \CP\ parities. In this 
case, the value of the dilution factor $2J_{s1}/J_0$ is estimated to 
be large~\cite{Browder:1999ng},
similar to the decay $\Bz\to\Dstarp\Dstarm$. 

The situation is more complicated if intermediate resonances such as
$D^+_{sJ}$ are present.  In this case, 
the parameter $J_{s2}$ is non-zero and $J_c$ can be large.
The resonant components are expected to be dominated by two $P$-wave excited 
$D_{s1}$ states~\cite{Browder:1999ng}. One such state is $D^+_{s1}(2536)$ that
 has a narrow width and does not contribute much to $J_{s2}$. It 
can be easily removed by imposing a mass window
requirement. The other $D^+_{s1}$ resonant state is predicted
in the quark model~\cite{God} to have a mass above the $\Dstarp\KS$ mass 
threshold with a large 
width. In this case, the $J_{s2}$ can be large. Therefore 
by studying the time-dependent asymmetry of $\Bz\to\Dstarp\Dstarm\KS$ in 
two different Dalitz regions, the sign of $\cos2\beta$ 
can be determined for a sufficiently large data set using the method 
described in Refs.~\cite{Browder:1999ng,Charles:1998vf,Colangelo:1999ny}.
This would allow the resolution of the $\beta\to\pi/2-\beta$ ambiguity
despite the large theoretical uncertainty of $2J_{s2}/J_{0}$. 
However, if the unknown $P$-wave $D^+_{s1}$ is 
the newly discovered $D^+_{sJ}(2317)$ or $D^+_{sJ}(2460)$,
both of which lie below
the $\Dstarp\KS$ mass threshold, then it will not contribute to
the decay $\Bztodstdstks$.
As a result, the time dependent analysis of $\Bz\to\Dstarp\Dstarm\KS$ not
only has a potential to measure the sign of \ctwob, but also can help
us to understand the possible structure of the excited charm meson spectrum.
  
In this paper,  
we present an improved measurement of the branching fraction of 
the decay $\Bztodstdstks$~\cite{Aubert:2003jq} and a search 
for intermediate resonant decays.
We also perform a time-dependent \CP asymmetry analysis 
to study the possible resonant contributions
and the sign of \ctwob .

The data used in this analysis comprise
$(230\pm 2)$ million \upsbb decays collected with the
\babar\ detector at the PEP-II storage rings. The \babar\ detector
is described in detail elsewhere~\cite{Aubert:2001tu}. 
We use a Monte Carlo (MC) simulation 
based on GEANT4~\cite{Agostinelli:2002hh}
to validate the analysis procedure and 
to study the relevant backgrounds.

We select $\Bztodstdstks$ decays by combining two oppositely 
charged $D^{*}$ candidates reconstructed in the modes
$\Dstarp\to\Dz\pip$ and $\Dstarp\to\Dp\piz$ 
with a \KS candidate.
We include the $\Dstarp\Dstarm$ combinations
$(\Dz\pip, \Dzb\pim)$ and $(\Dz\pip, \Dm\piz)$,
but not $(\Dp\piz,\Dm\piz)$ because of
the small branching fraction and large backgrounds.
To suppress the $\epem\to\qqbar \;(q=u,d,s,\,{\rm{and}}\; c)$
continuum background, we require the ratio of the
second and zeroth order Fox-Wolfram moments~\cite{Fox:1978vu} to be
less than 0.5. 

Candidates for \Dz and \Dp mesons are reconstructed in the modes
$\Dz\to\Km\pip$, $\Km\pip\piz$, $\Km\pip\pip\pim$, 
and  $\Dp\to\Km\pip\pip$, by selecting track combinations with invariant
mass within $\pm 2\,\sigma$ of the nominal $D$ masses~\cite{Eidelman:2004wy}.
The resolution $\sigma$ is measured using a large data sample of inclusive $D$ 
decays. It is equal to
7.0\,\mevcc for $\Dz\to K^-\pi^+$ decays, 13.5\,\mevcc for
$\Dz\to K^-\pi^+\pi^0$ decays, 5.7\,\mevcc for
$\Dz\to K^-\pi^+\pi^-\pi^+$ decays, and
5.6\,\mevcc for $\Dp\to K^-\pi^+\pi^+$ decays.

The \KS candidates are reconstructed from two oppositely-charged
tracks with an invariant mass within 
15\,\mevcc of the nominal \KS mass~\cite{Eidelman:2004wy},
which is equivalent to slightly less than $5\,\sigma$ of the measured
\KS mass resolution. The $\chi^2$ probability of 
the $\pip\pim$ vertex fit must be greater than $0.1\,\%$.
To reduce combinatorial background, we require the measured proper
decay time of the \KS to be greater than 3 times its uncertainty. 
Charged kaon candidates, except for the one in the decay $\Dz\to K^-\pi^+$,
are required to be inconsistent with the pion hypothesis, as 
inferred from the Cherenkov angle measured by the Cherenkov 
detector and the ionization energy loss measured by the 
charged-particle tracking system~\cite{Aubert:2001tu}. 
Neutral pion candidates are formed
from pairs of photons detected in the electromagnetic calorimeter~\cite{Aubert:2001tu},
each with energy above 30\,\mev. The mass of the pair must be within
30\,\mevcc of the nominal \piz mass, and their summed energy 
is required to be greater than 200\,\mev. In addition,
a mass-constrained fit is applied to the \piz candidate.

The \Dz and \Dp candidates are subject to a mass-constrained fit
prior to the formation of the \Dstarp candidates. 
The slow \pip from the \Dstarp decay is required to have a momentum in the 
\FourS center-of-mass (CM) frame less than 450\,\mevc.
The slow \piz from the \Dstarp must have a momentum between 
$70$ and $450\,\mevc$ in the CM frame.
No requirement on the photon-energy sum is applied to the
\piz candidates from the \Dstarp decays. The  
$\Dstarp$ mass is required to be within 4\,\mevcc of the 
nominal $\Dstarp$ mass, corresponding to slightly 
more than $3\,\sigma$ of the measured $\Dstarp$ mass resolution.

For each $\Bztodstdstks$ candidate, we calculate the difference of 
the \Bz candidate energy $E^*_B$ from 
the beam energy $E^{*}_{\rm{Beam}}$, 
$\Delta E\equiv E^*_B-E^{*}_{\rm{Beam}}$, 
in the CM frame.
In order to reduce the combinatorial background further,  
$|\Delta E|$ is required to be 
less than $25\,\mev$, which is equivalent to
$2.5\,\sigma$ of the measured $\Delta E$ resolution.

\begin{figure*}[t]
\begin{center}
\scalebox{0.8}{\includegraphics{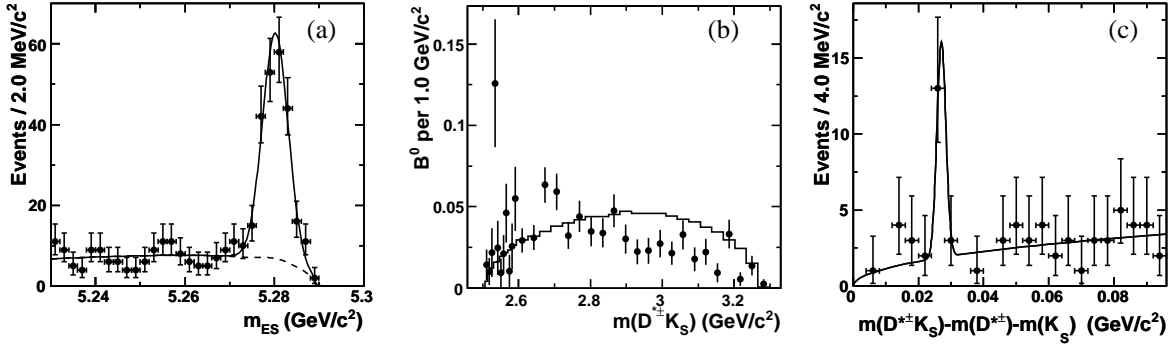}}
\caption{ (a) Measured distribution of $m_{\rm ES}$.
The solid line is the projection of the fit result.
The dashed line represents the background components.
(b) The efficiency-corrected yield of $\Bztodstdstks$ 
signal events as a function of $m(\Dstarpm\KS)$  
in data (points) and in three-body phase-space
signal MC (histogram) with an arbitrary normalization. Errors shown 
are statistical only. Note that the vertical axis shows
events per unit $m(\Dstarpm\KS)$, not the events in each 
bin.
(c) Measured distribution of $m(\Dstarpm\KS)-m(\Dstarpm)-m(\KS)$
in the region $\mes > 5.27\,\gevcc$.  
The solid line is the projection of the fit result.
}
\label{fig::Br}
\end{center}
\end{figure*}

The beam energy-substituted mass, 
$m_{\rm{ES}}=\sqrt{E^{*2}_{\rm{Beam}}-p^{*2}_B}$, 
where $p^*_B$ is
the $B^0$ candidate momentum in the CM frame, is used
to extract the signal yield from the events satisfying the 
aforementioned selection. We select \Bz
candidates with $m_{\rm{ES}}\ge5.23\,\gevcc$. 
On average we have $1.25$ \Bz candidates per event. If more than one 
candidate is selected in an event, we retain the one
with the smallest $|\Delta E|$. 
Studies using MC samples show that this procedure results in the 
selection of the correct \Bz candidate more than 95\,\% of the time.

The total probability density function (PDF) is 
the sum of the signal and background components.
The signal $m_{\rm{ES}}$ PDF is modeled
by a Gaussian function and the combinatorial background is described by 
an ARGUS~\cite{Albrecht:1990cs} function. MC studies show that there is a
small peaking background from $\Bp\to\Dstarzb\Dstarp\KS$ in which a 
$\Dzb$ originating from a \Dstarzb\ decay is combined with a random soft \pim\ to
form a \Dstarm\ candidate. The peaking background is described by 
the same PDF as the
signal, its fraction with respect to the signal yield is fixed to be 
1.4\,\%, determined from the MC simulation. 
An unbinned maximum likelihood (ML) fit to the $m_{\rm{ES}}$ distribution yields
$201\pm17\stat$  signal events, where the mean and width of the signal 
Gaussian, as well as the ARGUS shape parameters are allowed to float in 
the fit. In the region of $m_{\rm{ES}}> 5.27\,\gevcc$,
the signal purity is approximately 79\,\%. The fit result is shown 
in Fig.~1a.

To correct for variations in signal efficiency across the 
$\Dstarp\Dstarm\KS$ Dalitz plane, we calculate the branching fraction using
the {\it sPlots} method~\cite{Pivk:2004ty}:
\begin{equation}
{\cal B} = \sum_{i} \frac{w_{\rm sig}(m_{\rm ES,i})}{N_{\BB} \cdot
\epsilon_i \cdot {\cal B}_{\rm sub}},
\end{equation}
where the sum is over all events $i$,  
$\epsilon_i$ is the efficiency estimated from the simulated events in 
the vicinity of each data point in the Dalitz plane, 
$\mathcal{B}_{\rm sub}$ is the product of the branching fractions of the
sub-decays, and $w_{\rm sig}$ is an event-dependent
signal weight that is defined as~\cite{Pivk:2004ty}:
\begin{equation}
w_{\rm sig}(m_{\rm ES,i})=\frac{\sum^{N_s}_{j=1}V_{{\rm sig},j}P_{j}(m_{\rm ES,i})}
{\sum^{N_s}_{j=1}N_j P_{j}(m_{\rm ES,i})},
\end{equation}
which is calculated from the yield $N_j$ of the $j$-th PDF component 
$P_j$ in the fit, and the covariance matrix elements $V_{{\rm sig},j}$ between 
the signal yield $N_{\rm sig}$ and $N_j$. The $N_s$ is the number of 
PDF components in the fit.

We investigate the production of intermediate resonances
by examining the invariant  mass distribution
of the $\Dstarpm$ and $\KS$ combinations. Fig.~1b shows the 
projected distribution 
of $m(\Dstarpm\KS)$ from $\Bz\to\Dstarp\Dstarm\KS$ signal events
after efficiency correction using the {\it sPlots} technique.
A peak is seen at the value of $D^+_{s1}(2536)$ mass. We
do not observe evidence of the $D^+_{s2}(2573)$. 
The events tend to cluster toward lower values of 
$m(\Dstarpm\KS)$ (below about $2.9\,\gevcc$),
in contrast to the phase space model, as shown in Fig.~1b.

To extract the signal yield of $\Bz\to\Dstarm D^+_{s1}(2536)$, 
we perform an unbinned ML 
fit to the  $\Delta m= m(\Dstarpm\KS)-m(\Dstarpm)-m(\KS)$ distribution 
in the region $m_{\rm ES} >5.27\,\gevcc$ with a PDF given by the sum of a Gaussian
shape for the signal and a threshold function $\Delta m^a\exp(b\Delta m)$
for the background. The mean and width of the signal Gaussian, as well as 
the background PDF parameters $a$ and $b$ are allowed to float in the fit.
The fit yields $12.3\pm 4.0\stat$ signal events, as shown in Fig.~1c. 
The significance is estimated to be $4.6\,\sigma$ using the 
log-likelihood ratio between a fit with signal and another with none.
The significance is dominated by the statistical uncertainty and has
little contribution from the systematic uncertainty corresponding to 
the estimate of the signal yield.
The fitted signal mean and width are consistent with the MC simulation.
We repeat the fit in different $\Delta m$
regions up to the kinematic limit, as well as using different background
parameterizations. All of these give consistent signal yields of 
$D^+_{s1}(2536)$. We also examine the
$\Delta m$ distribution in the $m_{\rm ES}\le 5.27\,\gevcc$ region, and see no
peaking structure.

The systematic uncertainties of the branching fraction measurements are dominated
by the uncertainty of the charged track reconstruction efficiency ($10.7\,\%$). 
Other sources also contribute to the systematic errors, such as the kaon particle
identification efficiency ($3.9\,\%$), $\piz$ reconstruction efficiency ($3.5\,\%$), 
branching fractions of the $D$ decays ($5.8\,\%$), 
determination of the number of \BB\ in the
data sample ($1.1\,\%$), event selection criteria ($5.0\,\%$),
and the estimate of the peaking background fraction ($1.8\,\%$).
The measured branching fraction is:
$$
\mathcal{B}(\Bztodstdstks)= 
(4.4\pm0.4\pm0.7)\times 10^{-3},
$$
where the first  uncertainty is the statistical and
the second is systematic. 
Our result is in good agreement 
with the previous \babar\ measurement~\cite{Aubert:2003jq}. 
We also measure the intermediate resonant decay branching fraction
and find:
$$
\begin{array}{c}
\mathcal{B}(\Bz\to \Dstarm D^+_{s1}(2536)) \times
\mathcal{B}(D^+_{s1}(2536)\to \Dstarp\KS)  \\ \\
= (4.1\pm1.3\pm0.6)\times 10^{-4}.
\end{array}
$$
The fraction of the decay
$\Bztodstdstks$ through the intermediate $D^+_{s1}(2536)$
resonance is measured to be $0.092\pm0.024\stat\pm0.001\syst$.

We subsequently perform a time-dependent analysis using 
the event sample described previously. In the time-dependent analysis,
we require that the invariant mass
of the $\Dstarpm$ and $\KS$ combination be larger than
$2.55\,\gevcc$ in order to reject the narrow $D^+_{s1}(2536)$ resonant 
decays.

For the time-dependent \CP analysis, we use information from 
the other $B$ meson in the event to tag the initial flavor of the
fully reconstructed $\Bztodstdstks$ candidate. The 
decay rate $f_+ (f_-)$ for a neutral $B$ meson accompanied by  a $B^{0}
(\Bzb)$ tag is given by
\begin{widetext}
\begin{equation}
%\begin{array}{rcl}
f_\pm(\deltat) \propto 
{\rm e}^{ - | \deltat |/\tau_{B^0} }
\left\{ (1\mp\Delta\omega) \pm (1-2\omega)\times  
 \left[\eta_y\frac{J_c}{J_0}\cos{ (\deltamd  \deltat)} -
 \left(\frac{2J_{s1}}{J_0}\stwob +\eta_y\frac{2J_{s2}}{J_0}\ctwob\right)
\sin{ (\deltamd  \deltat) }  \right]  \right\},
 \label{eq:CP}
%\end{array}
\end{equation}
\end{widetext}
where $\Delta t = t_{\rm rec} - t_{\rm tag}$ is the difference between
the proper decay time of the reconstructed signal $B$ meson 
($B_{\rm rec}$) and
that of the tagging $B$ meson ($B_{\rm tag}$),
$\tau_{\Bz}$ is the \Bz lifetime, and \deltamd is the mass difference
determined from the \Bz-\Bzb oscillation frequency~\cite{Eidelman:2004wy}.
The average mistag probability $\omega$ describes
the effect of incorrect tags, and $\Delta\omega$
is the difference between the mistag rate for $\Bz$ and $\Bzb$. 

\begin{figure}[tb]
\begin{center}
\scalebox{0.45}{\includegraphics{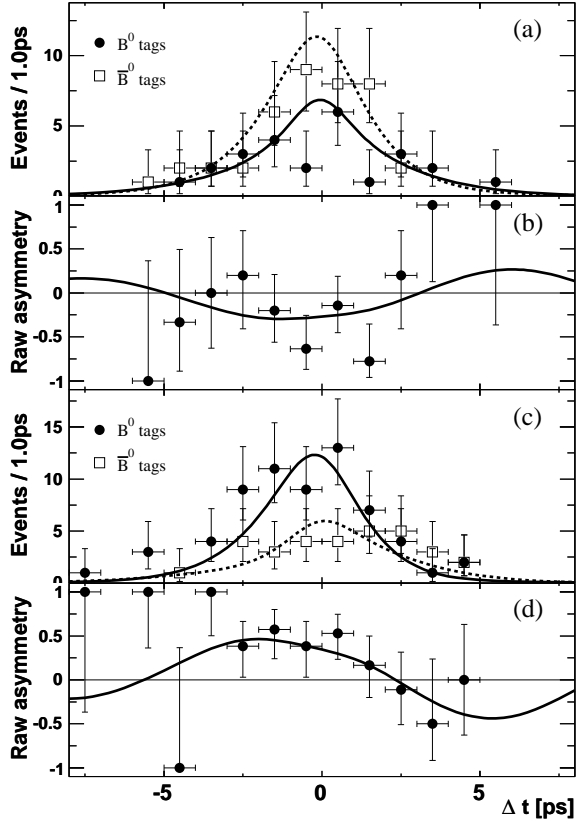}}
\caption{ 
(a) The distribution of $\Delta t$ in the region
$\mes > 5.27\,\gevcc$ for \Bz (\Bzb) tag candidates in the
half Dalitz space $s^+<s^-$ ($\eta_y=-1$).
The solid (dashed) curve represents the fit projections in $\Delta t$ for
\Bz (\Bzb) tags. (b)
The raw asymmetry
$(N_{\Bz}-N_{\Bzb})/(N_{\Bz}+N_{\Bzb})$, as functions of \deltat , where
$N_{\Bz}$ ($N_{\Bzb}$) is the number of candidate with \Bz (\Bzb) tag.
(c) and (d) contain the corresponding information for the
\Bz candidates in the other half Dalitz space $s^+>s^-$ ($\eta_y=+1$).}
\label{fig::CP}
\end{center}
\end{figure}

The technique used to measure the \CP asymmetry is analogous to that 
used in previous
\babar\ measurements as described in Ref.~\cite{Aubert:2002rg,Aubert:2004zt}.
We calculate the time interval \deltat between the two $B$ decays from the measured 
separation $\Delta z$ between the decay vertices of $B_{\rm rec}$ and $B_{\rm tag}$ 
along the collision ($z$) axis~\cite{Aubert:2002rg}. The $z$ position of the 
$B_{\rm rec}$ vertex is determined from the charged daughter tracks. The $B_{\rm tag}$ 
decay vertex is determined by fitting charged tracks not belonging to the $B_{\rm rec}$ 
candidate to a common vertex, employing constraints from the beam spot
location and the $B_{\rm rec}$ momentum~\cite{Aubert:2002rg}. 
Only events with a $\Delta t$ uncertainty less than $2.5\,\mbox{ps}$ and a measured
$|\Delta t|$ less than $20\,\mbox{ps}$ are accepted.
We perform a simultaneous unbinned maximum likelihood fit to
the $\Delta t$ and $m_{\rm ES}$ distributions
to extract the \CP asymmetry. The signal PDF in $\Delta t$ 
is given by Eq.~\ref{eq:CP} convolved with an empirical 
$\Delta t$ resolution function~\cite{Aubert:2002rg}.
Both the signal mistag probability and $\Delta t$ resolution function are determined
from a sample of neutral $B$ decays to flavor eigenstates, $B_{\rm flav}$.

The background $\Delta t$ distributions are parameterized with an empirical
description that includes zero and 
non-zero lifetime components~\cite{Aubert:2002rg}.
We also allow the non-zero lifetime
background to have effective \CP asymmetries and let them float 
in the likelihood fit.

The fits to the data yield
\begin{eqnarray}
\frac{J_c}{J_0} &=& 0.76\pm 0.18\stat\pm 0.07\syst \nonumber \\
\frac{2J_{s1}}{J_0}\stwob &=& 0.10\pm 0.24\stat\pm 0.06\syst \\
\frac{2J_{s2}}{J_0}\ctwob &=& 0.38\pm 0.24\stat\pm 0.05\syst \nonumber\\\nonumber
\end{eqnarray}
Fig.~\ref{fig::CP} shows the $\Delta t$ distributions
and asymmetries in yields between $B^0$ and $\Bzb$
tags, overlaid with the projection of the likelihood fit result.
The effective \CP asymmetries in the background are found to be
consistent with zero within statistical uncertainties.
As a cross check, we also repeat the fit by allowing the \Bz lifetime to 
float. The obtained \Bz lifetime is in a good agreement with
its world average~\cite{Eidelman:2004wy} 
within the statistical uncertainty.

The sources and estimates of systematic uncertainties are 
summarized in Table~\ref{tab:systematics}.
Since the signal reconstruction efficiency is not uniform over the entire 
Dalitz space, the different \CP components may not have the same acceptance. Therefore
the measured parameters will deviate slightly from their true values. 
We estimate the possible bias using the signal MC weighted according to the 
expected theoretical Dalitz distributions in Ref.~\cite{Browder:1999ng}.
Because of the lack of knowledge of the unknown $D^+_{s1}$ 
state, we vary its mass and width over a 
wide range. The largest bias of the measured parameters 
${J_c}/{J_0}$, $(2J_{s1}/J_0)\stwob$ and $(2J_{s1}/J_0)\ctwob$
are taken as the corresponding systematic uncertainties on the
acceptance effect.

The other systematic uncertainties arise
from the possible backgrounds that tend to peak
under the signal and their
\CP asymmetries, the assumed parameterization of the
$\Delta t$ resolution function, the possible differences between the
$B_{\rm flav}$ and $\Bz\to\Dstarp\Dstarm\KS$ tagging performances, knowledge
of the event-by-event beam-spot position, and the possible
interference between  the suppressed
$\bar{b}\to\bar{u}c\bar{d}$ amplitude and the favored
$b\to c\bar{u}d$ amplitude for some tag-side decays~\cite{Long:2003wq}.
They also include the systematic uncertainties from the finite MC
sample used to verify the fitting method. All 
the systematic
uncertainties are found to be much smaller than the statistical
uncertainties.
\begin{table}[!htb]
\begin{ruledtabular}
\begin{tabular}{lccc}
Source  &  I & II & III 
\\ \colrule
Acceptance                    & 0.060 & 0.040 & 0.030  \\
Peaking backgrounds           & 0.009 & 0.016 & 0.002  \\
\deltat\ resolution function  & 0.015 & 0.006 & 0.008  \\
Mistag fraction differences   & 0.016 & 0.015 & 0.015  \\
Detector Alignment            & 0.005 & 0.015 & 0.015  \\
$\deltamd$, $\tau_B$          & 0.001 & 0.001 & 0.001  \\
MC statistics                 & 0.021 & 0.032 & 0.032  \\ 
Others                        & 0.005 & 0.004 & 0.005  \\ \hline
Total                         & 0.068 & 0.058 & 0.050  \\
\end{tabular}
\end{ruledtabular}
\caption{
Sources of systematic error on ${J_c}/{J_0}$ (column I), 
$(2J_{s1}/J_0)\stwob$ (column II)
and $(2J_{s1}/J_0)\ctwob$ (column III).}
\label{tab:systematics}
\end{table}

In summary, we have reported 
an improved branching fraction measurement of the decay $\Bztodstdstks$
that supersedes the previous \babar\ result~\cite{Aubert:2003jq}.
We also find evidence for the decay 
$\Bz\to\Dstarm D^+_{s1}(2536)$ with  $4.6\,\sigma$
significance. A time-dependent \CP asymmetry analysis has
also been performed. The measured 
$J_c/J_0$ is significantly different from zero, which may indicate
that there is a sizable resonant contribution to the 
decay $\Bz\to\Dstarp\Dstarm\KS$ from a unknown $D^+_{s1}$ state with large
width, according to Ref.~\cite{Browder:1999ng}. We measure that
$(2J_{s2}/J_0)\ctwob=0.38\pm0.24\stat\pm0.05\syst$. 
Under the assumption that there is a significant broad resonant contribution
to the decay $\Bz\to\Dstarp\Dstarm\KS$,
it implies that the sign of 
\ctwob is preferred to be positive at the 94\,\% confidence level
if the theoretical parameter $J_{s2}/J_0$ is positive, as predicted 
in Ref~\cite{Browder:1999ng}. 

We are grateful for the excellent luminosity and machine conditions
provided by our \pep2\ colleagues, 
and for the substantial dedicated effort from
the computing organizations that support \babar.
The collaborating institutions wish to thank 
SLAC for its support and kind hospitality. 
This work is supported by
DOE
and NSF (USA),
NSERC (Canada),
IHEP (China),
CEA and
CNRS-IN2P3
(France),
BMBF and DFG
(Germany),
INFN (Italy),
FOM (The Netherlands),
NFR (Norway),
MIST (Russia), and
PPARC (United Kingdom). 
Individuals have received support from CONACyT (Mexico), 
Marie Curie EIF (European Union),
the A.~P.~Sloan Foundation, 
the Research Corporation,
and the Alexander von Humboldt Foundation.


\begin{thebibliography}{99}
\bibitem{conjugate}
We imply charge conjugate modes throughout the paper.
\end{thebibliography}
\end{document}